\documentclass{aa}
\usepackage{epsfig,lscape}

\newcommand{\ergscm}{erg\,s$^{-1}$\,cm$^{-2}$}

\begin{document}
\thesaurus{02.12.3 ;11.09.1 Mkn 110; 11.19.1}
\title{Strong optical line variability in Mkn~110
\thanks{Visiting Astronomer, German-Spanish Astronomical Centre, Calar Alto, operated by the Max-Planck-Institute for Astronomy, Heidelberg, jointly with the Spanish National Commission for Astronomy.}
}
\author{K. Bischoff \and W. Kollatschny}
\institute{ 
 Universit\"{a}ts-Sternwarte G\"{o}ttingen, Geismarlandstra{\ss }e 11, 
D-37083 G\"{o}ttingen, Germany}
\offprints{W. Kollatschny}
\mail{kbischo@gwdg.de} 
\date{Received date; accepted date}
\maketitle
\begin{abstract}

We present results of a long-term variability campaign
on the Seyfert 1 galaxy Markarian 110.
Mkn~110 is a narrow-line Seyfert 1 object hosted in a morphological peculiar galaxy.
We monitored the optical continuum and the line intensities as well as their
profiles over a time interval of nearly ten years.
The continuum and the Balmer lines varied by a factor of 2 to 5
within two years. The HeII$\lambda$4686 line showed exceptional 
intensity variations of a factor of eight.
We detected an additional independent very broad-line region in
high intensity stages of the Balmer and HeII lines.
The CCF analysis of the HeII line indicates that this very broad-line region 
originates at a distance of 9 light days only from the central ionizing
source.
          
\end{abstract}
\section{Introduction}
Markarian 110 is a nearby (z=0.0355) Seyfert~1 galaxy with highly irregular morphology.
 The apparent magnitude of the total system
 is m$_V$=15.4~mag (Weedman 1973) corresponding to M$_V$=-20.4~mag 
($H_0$=75~km s$^{-1}$ Mpc$^{-1}$). A foreground star is
projected on the host galaxy at a distance
of 6 arcsec to the nucleus in north-east direction
(see Fig.~1). Therefore, it has been supposed 
in some early papers that
Mkn~110 might be a double nucleus galaxy (Petrosian et al.\ 1978).

On the other hand the peculiar morphology of Mkn~110 is an indication
for a recent interaction and/or merging event in this galaxy (Hutchings \& Craven 1988).
\begin{figure}
%epsfig{figure=m110rg+c2.cd.ps,width=8.5cm,clip=}
\epsfig{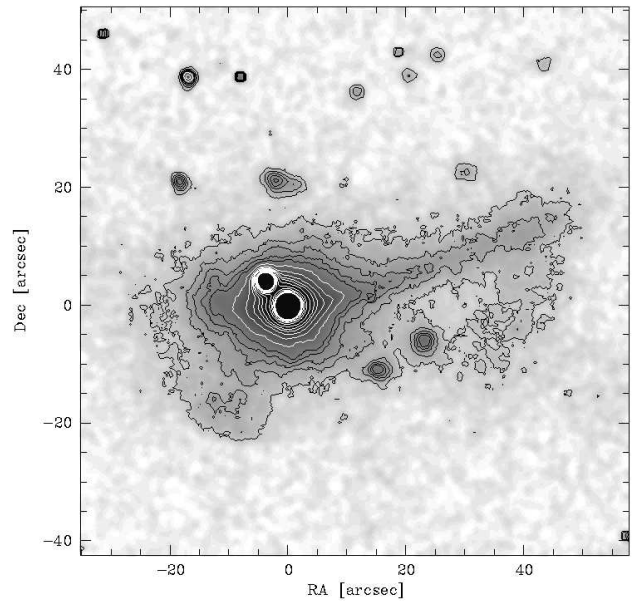}
\caption{Optical R-band image of Mkn~110. North is up, east is to the left.}
\end{figure}
One can clearly recognize a tidal arm to the west 
with a projected length of 50 arcsec (corresponding to 35 kpc) and further signs of asymmetry in the disturbed host galaxy on the R-band image (Fig.~1).

Ten years ago we started a long-term variability campaign to study the
continuum and emission line intensity variations in
selected AGNs. Besides our principal interest in the long-term
variability behaviour of these galaxies
on its own we want to compare the individual variability
properties with those of other galaxies from 
the international AGN watch campaign (Peterson et al.\ 1991) (e.g.\ NGC~5548, Kollatschny \& Dietrich 1996) and
LAG campaign (Robinson 1994) (e.g.\ NGC~4593, Kollatschny \& Dietrich 1997). 
Further results on
 continuum and H$\beta$ variations in Mkn~110 have been published in a recent
paper on variability of Seyfert 1 galaxies (Peterson et al.\ 1998a).    

\section{Observations and data reduction}
We took optical spectra of Mkn~110 at 24 epochs from February 1987 until
June 1995. The sampling of the observations extends from days to years.
 In Table 1 we list our observing dates and the
corresponding Julian Dates. The spectra were obtained at Calar Alto Observatory in Spain with the 2.2~m and 3.5~m telescopes 
 as well as at McDonald Observatory in Texas with the 2.1~m and 2.7~m telescopes.
The individual exposure times range from 10 minutes to 1 hour (see Table~1).
We used spectrograph slits with projected widths 
of 2 to 2.5 arcsec and  2 arcmin length
under typical seeing conditions of 1 to 2 arcsec.
We extracted spectra of the central 5 arcsec.
The slit was oriented in north-south direction in most cases.

To investigate the spatial extension of the narrow line region we took spectra
at  different position angles:
0$^{\circ}$, 45$^{\circ}$, 90$^{\circ}$, and 135$^{\circ}$.
A possible extended [OIII]$\lambda$5007 emission line flux was always
less than 3\% of the nuclear point-like emission.
Mkn~110 has been inspected by Nelson et al.\ (1996)
with the Hubble Space Telescope WFPC-1 in the near infrared spectral range.
They detected a dominant unresolved nucleus in this galaxy.

Our optical spectra typically cover a wavelength range from 4000\,\AA\ 
to 7200\,\AA\ with a spectral resolution of 3 to 7\,\AA\ per pixel.
We used different CCD detectors in the course of
this monitoring program: until 1989 a RCA-chip (1024x640), in  January and July 1992 a GEC-chip (1155x768), and in August 1992 a Tektronix-chip (1024x1024).

The reduction of the spectra (flat fielding, wavelength calibration,
night sky subtraction, flux calibration, etc.) was done in a homogeneous way using the ESO MIDAS package.

The absolute calibration of our spectra was achieved by scaling the [OIII]$\lambda$5007 line of all spectra to those obtained
under photometric conditions. Our absolute [OIII]$\lambda$5007 flux corresponds
 within 5\% to that obtained by Peterson et al.\ (1998a). For a better comparison of these two data sets we will use exactly the same [OIII]$\lambda$5007 flux of 2.26 $10^{-13}$ \ergscm. Furthermore, we corrected all our data for small spectral shifts and resolution differences with respect to a mean reference spectrum using an automatic scaling program of van Groningen \& Wanders (1992).

\begin{table}
\caption{Log of observations}
\begin{tabular}{cccccccccc}
\hline 
\noalign{\smallskip}
Julian Date & UT Date & Telescope & Exp. time \\
2\,440\,000+&         &           &  [sec.]   \\
(1) & (2) & (3) & (4) \\ %& (5) & (6) & (7) & (8) & (9) \\ 
\noalign{\smallskip}
\hline 
\noalign{\smallskip}
6828 & 1987-02-01 & CA 3.5  & 1500 \\
7229 & 1988-03-08 & CA 3.5  & 4800 \\
7438 & 1988-10-03 & CA 3.5  & 1200 \\
7574 & 1989-02-16 & CA 2.2  & 3000 \\
7663 & 1989-05-16 & CA 3.5  & 1800 \\
7828 & 1989-10-28 & CA 2.2  & 1200 \\
8632 & 1992-01-09 & CA 3.5  &  600 \\
8812 & 1992-07-08 & CA 3.5  & 1800 \\
8860 & 1992-08-25 & CA 2.2  &  900 \\
8862 & 1992-08-27 & CA 2.2  &  900 \\
8864 & 1992-08-25 & CA 2.2  &  900 \\
9078 & 1993-03-31 & CA 2.2  & 3600 \\ 
9080 & 1993-04-02 & CA 2.2  & 3600 \\
9083 & 1993-04-05 & CA 2.2  & 2400 \\
9123 & 1993-05-15 & MDO 2.1 & 1800 \\
9237 & 1993-09-06 & CA 3.5  & 1200 \\
9419 & 1994-03-07 & MDO 2.1 & 2100 \\
9595 & 1994-08-30 & CA 2.2  & 1200 \\
9776 & 1995-02-27 & MDO 2.1 & 1800 \\
9785 & 1995-03-08 & MDO 2.1 & 1200 \\
9786 & 1995-03-09 & MDO 2.1 & 1800 \\
9787 & 1995-03-10 & MDO 2.1 & 1200 \\
9811 & 1995-04-03 & MDO 2.1 & 1200 \\
9870 & 1995-06-01 & MDO 2.7 &  900 \\
\noalign{\smallskip}
\hline 
\end{tabular}
%\noalign{\smallskip}

CA = Calar Alto Observatory\\
MDO = McDonald Observatory
\end{table}

The R-band image of Mkn~110 (Fig.~1) was taken 
with the 2.2m telescope at Calar Alto observatory
on September 20, 1993 with an exposure time of 6 minutes.
Again, we reduced this CCD image with the ESO MIDAS package.

In the course of our discussion we additionally will make use of archival 
IUE spectra of Mkn~110 taken on February 28 and 29, 1988.

\section{Results}
Some typical spectra are plotted in Fig.~2 showing the range of intensity variations.
\begin{figure}
%\picplace{8cm}
%centerline{\psfig{figure=overspek.ps,width=6.5cm,angle=270,clip=}}
\centerline{\psfig{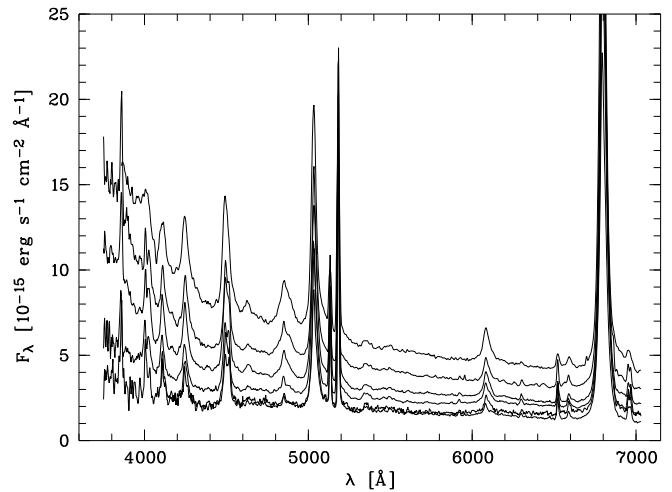}}
\caption{Normalized spectra of Mkn~110 taken at different epochs in
Oct. 1988, Oct. 1989, March 1988, Feb.
 1989, May 1989, Jan. 1992 (from bottom to top).}
\end{figure}
Immediately one can recognize the strong variations in the continuum,
in the Balmer lines and especially in the HeII$\lambda$4686 line. The 
continuum variations are most pronounced
 in the blue section. The continuum
gradient changes as a function of intensity.
The emission line profiles of the Balmer lines in Mkn~110 are quite narrow
 (FWHM(H$\beta$)=1800~km~s$^{-1}$) similar to those of
the so called narrow-line Seyfert 1 galaxies.
Week FeII emission is present in the spectra
 blending the red line wings of [OIII]$\lambda$5007 and H$\beta$. 
We measured the integrated
 intensity of FeII line blends between 5134\,\AA\ and 5215\,\AA. 
The main FeII components in this region are the 5169\,\AA\
and 5198\,\AA\ lines belonging to the multiplets 42 and 49.
The FeII line flux (2.0 10$^{-14}$ \ergscm\ )
remained constant during our variability campaign within the error of 10\%.

Difference spectra with respect to our minimum stage in October 1988
are plotted in Fig.~3. All narrow line components cancel out.
The FeII lines disappear in the difference spectra as well.
\begin{figure}
%\picplace{8cm}
%centerline{\psfig{figure=overdiff.ps,width=6.5cm,angle=270,clip=}}
\centerline{\psfig{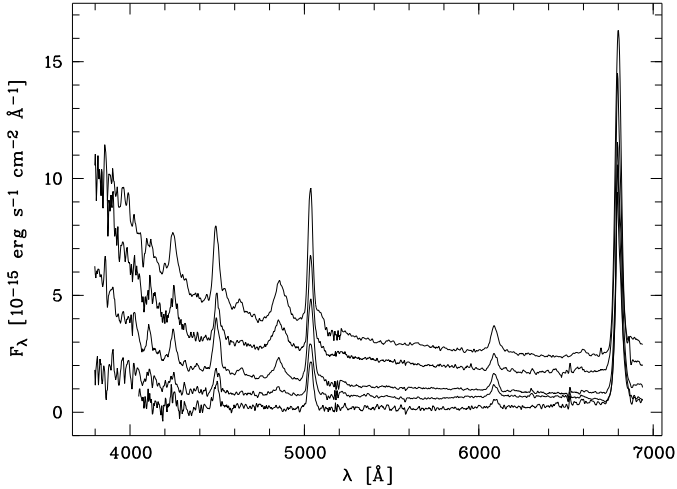}}
\caption{Difference spectra with respect to our minimum stage in Oct. 1988;
otherwise same epochs as in Fig.~2.}
\end{figure}
In the Balmer profiles (e.g.\ H$\beta$) very broad, slightly redshifted 
components stand out in the high intensity stages.
\subsection{Line and continuum variations}
\noindent
The results of our continuum intensity measurements
at 3750\,\AA, 4265\,\AA, and 5100\,\AA\ as well as 
the integrated line intensities of
H$\alpha$, H$\beta$, HeII$\lambda$4686, HeI$\lambda$5876, and  
HeI$\lambda$4471 are given in Table~2.
 The individual light curves are plotted in Fig. 4.

\begin{figure*}
 \hbox{\psfig{figure=MS8273.f4a,width=56mm,height=88mm,angle=270}\hspace*{-2mm}
       \psfig{figure=MS8273.f4b,width=56mm,height=88mm,angle=270}}
 \hbox{\psfig{figure=MS8273.f4c,width=56mm,height=88mm,angle=270}\hspace*{-2mm}
       \psfig{figure=MS8273.f4d,width=56mm,height=88mm,angle=270}}
 \hbox{\psfig{figure=MS8273.f4e,width=56mm,height=88mm,angle=270}\hspace*{-2mm}
       \psfig{figure=MS8273.f4f,width=56mm,height=88mm,angle=270}}
 \hbox{\psfig{figure=MS8273.f4g,width=56mm,height=88mm,angle=270}\hspace*{-2mm}}
  \caption{Light curves of continuum flux at 5100\,\AA\ and 4265\,\AA\ (in units of 10$^{-15}$ erg cm$^{-2}$ s$^{-1}$\,\AA$^{-1}$) and integrated emission line flux of H$\alpha$, H$\beta$, HeII$\lambda$4686, HeI$\lambda$5876 and
 HeI$\lambda$4471 (in units of  10$^{-15}$ erg cm$^{-2}$ s$^{-1}$). The points are connected by a dotted line to aid the eye.}
\end{figure*}

The continuum intensities are mean values of the wavelength ranges given in Table 3, column (2). Line intensities were integrated in the listed limits after subtraction of a linear pseudo-continuum defined by the boundaries given in column (3). All wavelengths are given in the rest frame.

\setcounter{table}{2}
\begin{table}
\caption{Boundaries of mean continuum values and line integration limits}
\begin{tabular}{lccccccc}
\hline 
\noalign{\smallskip}
Cont./Line & Wavelength range & Pseudo-continuum \\
\noalign{\smallskip}
(1) & (2) & (3) \\
\noalign{\smallskip}
\hline 
\noalign{\smallskip}
Cont.~3750         & 3745\,\AA\ -- 3755\,\AA \\
Cont.~4265         & 4260\,\AA\ -- 4270\,\AA \\
Cont.~5100         & 5095\,\AA\ -- 5105\,\AA \\
HeII$\lambda 4686$ & 4600\,\AA\ -- 4790\,\AA & 4600\,\AA\ -- 4790\,\AA \\
HeI$\lambda 4471$  & 4430\,\AA\ -- 4530\,\AA & 4265\,\AA\ -- 4600\,\AA \\
HeI$\lambda 5876$  & 5800\,\AA\ -- 5960\,\AA & 4600\,\AA\ -- 5100\,\AA \\
H$\beta$           & 4790\,\AA\ -- 4935\,\AA & 4600\,\AA\ -- 5100\,\AA \\
H$\alpha$          & 6420\,\AA\ -- 6770\,\AA & 6420\,\AA\ -- 6770\,\AA \\
\noalign{\smallskip}
\hline \\
\end{tabular}
\end{table}

We started our monitoring program in 1987. Therefore,
our 5100\,\AA\ continuum light curve covers up
 a larger time interval than the light curve
of Peterson et al.\ (1998a) beginning in 1992.
The observing epochs are partly complementary in the common monitoring interval.
But, spectra taken  nearly simultaneously from both groups
(within one week) correspond with each other to better than 5\% in the continuum fluxes.
In Fig.~5 we compare our continuum light curve with that of Peterson
et al.\ (1988a) in the common interval of observations.
 Both light curves are in
very good accordance regarding to the intensity variations. Fig.~5 shows that
the data sets are not significantly undersampled.
\begin{figure}
%centerline{\psfig{figure=lccopetwk.ps,width=6.5cm,angle=270,clip=}}
\centerline{\psfig{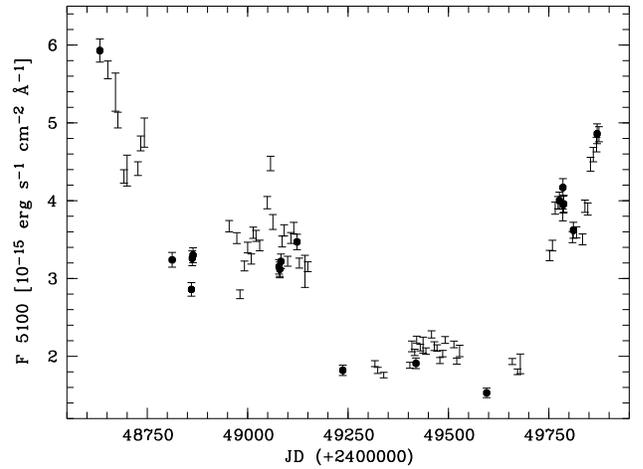}}
\caption{Comparison of our continuum light curve at 5100\,\AA\ (filled circles)
with that of Peterson et al.\ (1988a) between JD~2448600 and JD~2500000.}
\end{figure}
 However, our H$\beta$ intensities are systematically higher than those of Peterson et al.\ (1998a)  as we integrated over a larger wavelength range and 
we carried out a slightly different continuum subtraction.
This method led to a lower pseudo-continuum flux at 4790\,\AA.
The H$\beta$ fluxes are in perfect agreement if we multiply
the values given by Peterson et al.\ (1998a) by a factor of 1.15.

The pattern of the  continuum light curves at 5100\,\AA\ and 4265\,\AA\ (Fig.~4) is identical apart from their different
amplitudes. The HeII$\lambda$4686 light curve follows closely these 
continuum light curves. The light curves of the Balmer lines H$\alpha$ and
H$\beta$ are similar among themselves and the light curves of both HeI lines as well.

\begin{table}
\caption{Variability statistics}
\begin{tabular}{lccccccc}
\hline 
\noalign{\smallskip}
Cont./Line & F$_{min}$ & F$_{max}$ & R$_{max}$ & $<$F$>$ & $\sigma_F$ & F$_{var}$ \\
\noalign{\smallskip}
(1) & (2) & (3) & (4) & (5) & (6) & (7) \\ %(8) & (9) \\ 
\noalign{\smallskip}
\hline 
\noalign{\smallskip}
Cont.~3750         & 2.94 & 15.10 & 5.14 & 7.51  & 4.322 & 0.573 \\
Cont.~4265         & 1.96 &  7.56 & 3.86 & 3.91  & 1.533 & 0.391 \\
Cont.~5100         & 1.53 &  5.93 & 3.88 & 3.26  & 1.044 & 0.318 \\
HeII$\lambda 4686$ & 30.6 & 237.6 & 7.76 & 102.6 & 48.95 & 0.466 \\
HeI$\lambda 4471$  & 12.0 &  41.7 & 3.47 & 28.56 &  7.19 & 0.180 \\
HeI$\lambda 5876$  & 33.1 & 103.2 & 3.12 & 69.44 & 17.31 & 0.236 \\
H$\beta$           & 266  & 624.5 & 2.34 & 419.2 & 79.94 & 0.189 \\
H$\alpha$          & 876  & 2091  & 2.39 & 1572  & 265.3 & 0.168 \\
\noalign{\smallskip}
\hline 
%\noalign{\smallskip}
\end{tabular}

Continuum flux in units of
 10$^{-15}$\,erg\,sec$^{-1}$\,cm$^{-2}$\,\AA$^{-1}$.\\
Line flux in units of 10$^{-15}$\,erg\,sec$^{-1}$\,cm$^{-2}.$
\end{table}

Statistics of our measured continuum and emission line variations are presented in Table 4. We list minimum and maximum fluxes F$_{min}$ and F$_{max}$,
peak-to-peak amplitudes R$_{max}$ = F$_{max}$/F$_{min}$, the mean flux
over the entire period of observations $<$F$>$, the standard deviation $\sigma_F$, and the fractional variation 
\[ F_{var} = \frac{\sqrt{{\sigma_F}^2 - \Delta^2}}{<F>} \] 
as defined by Rodr\'iguez-Pascual et al.\ (1997).
The extreme variability amplitudes of Mkn~110 attract attention
compared to other galaxies, e.g. NGC~4593 (Dietrich,
Kollatschny et al.\ 1994) and the Seyfert~1 galaxies from the sample of Peterson et al.\ (1998a).

The variability amplitudes of the continuum increase towards the 
short wavelength region. These amplitudes as well as
 those of the emission line intensities
are exceptionally high. The variability amplitude of the HeII$\lambda$4686 line is unique compared to the other emission lines in Mkn~110
and compared to optical lines in other Seyfert galaxies.
In Fig.~6 we plot the line intensity ratios HeII$\lambda$4686/H$\beta$ and 
HeI$\lambda$5876/H$\beta$ as a function of continuum intensity at 5100\,\AA.
These line intensity ratios increase slightly for the HeI line but strongly for the highly ionized HeII line.

\begin{figure}
%centerline{\psfig{figure=hehb5100.ps,width=7cm,angle=270,clip=}}
\centerline{\psfig{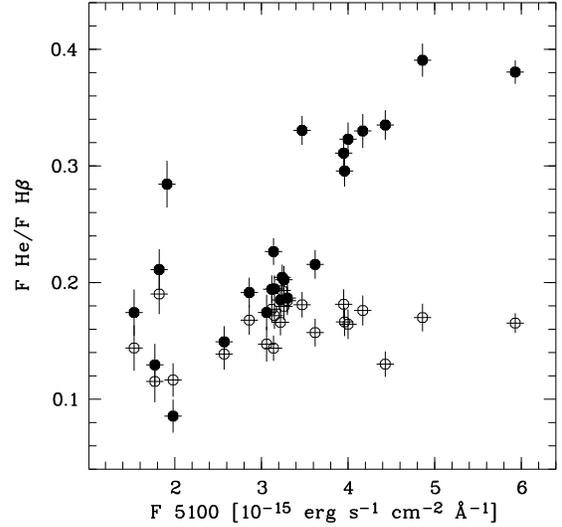}}
\caption{Line intensity ratios HeII$\lambda$4686/H$\beta$ (filled circles)
and HeI$\lambda$5876/H$\beta$ (open circle) as a function of continuum flux at 5100\,\AA.}
\end{figure}

\subsection{Balmer decrement}
We calculated Balmer decrement H$\alpha$/H$\beta$ values in the range from 3.2 to 4.3.
Simple photoionization calculations (Case B) result in a value of 2.8
for this line ratio (Osterbrock 1989).
Deviations of the observed Balmer decrement from the theoretical
value are often explained by wavelength dependent dust absorption
and/or by collisional excitation effects.
 We will show later on that the observed difference can not be explained
by dust absorption alone in the broad-line region clouds of Mkn~110.
There is a clear anti-correlation of the Balmer decrement
 with the continuum flux (Fig.~7).
\begin{figure}
%centerline{\psfig{figure=bd5100b.ps,width=7cm,angle=270,clip=}}
\centerline{\psfig{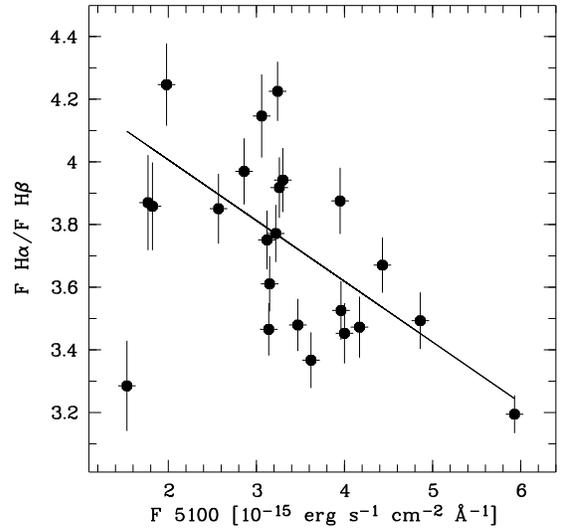}}
\caption{The Balmer decrement H$\alpha$/H$\beta$ as a function of
 continuum flux at 5100\,\AA. The minimum value of F 5100 was omitted for the linear fit.}
\end{figure}
One has to keep in mind that the individual Balmer lines and the
continuum of each spectrum originate in distinct regions of the BLR
at different times.
This will be confirmed by the cross-correlation analysis later on.
A very tight correlation can not be expected because of the
short-term variations in this galaxy.
The solid line in Fig.~7 is a linear fit to all our data points except for
the lowest continuum intensity point.
 At this epoch (JD +5959) the H$\alpha$ intensity was extreme low (Fig.~4)
in contrast to H$\beta$ and the continuum.
An anti-correlation of the Balmer decrement with the continuum flux has been
first noted in NGC~4151 by Antonucci \& Cohen (1983).

\subsection{UV spectra}
Two UV spectra haven been taken with the IUE satellite with a time interval
of one day only. These spectra have been taken nearly simultaneously (within 
8 days) to our
optical observations in March 1988. Therefore, these two spectra
are suitable for a determination of optical/UV line intensity ratios.
Fig.~8 shows an overplot of both short wavelength UV spectra taken with
an time interval of 1 day.
\begin{figure}
%centerline{\psfig{figure=uv.ps,width=6.5cm,angle=270,clip=}}
\centerline{\psfig{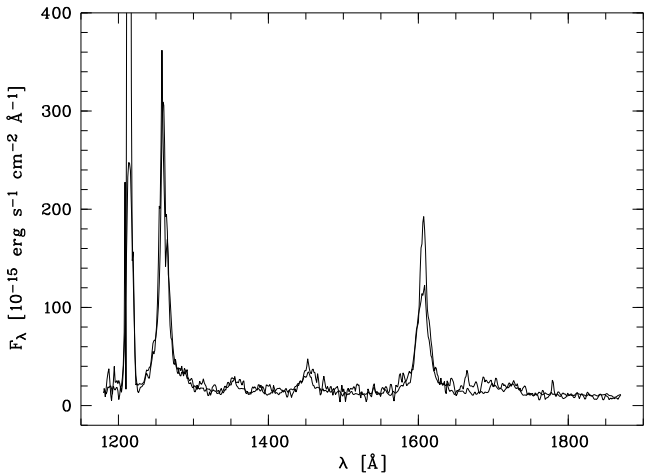}}
\caption{Short wavelength IUE UV spectra of Mkn~110 taken on Feb. 28
and Feb. 29, 1988} 
\end{figure}
 The spectra are identical in the continuum and in the emission lines 
 within the error limits. The different flux values in the center of the 
 CIV$\lambda$1550 line are due to saturation effects in one of the spectra.

We determined an integrated Ly$\alpha$ flux of  
$(4.2 \pm 0.3) 10^{-12}$ \ergscm. Comparison with the optical spectra results
in a Ly$\alpha$/H$\beta$ ratio of 11.0 at the observing epoch March, 1988.

The HeII$\lambda$1640 flux amounts to $(1.4 \pm 0.2) 10^{-13}$ \ergscm. The  
HeII$\lambda$1640/HeII$\lambda$4686 ratio of 2.45 is about a factor of two
lower than that of typical photoionization models 
but consistent with other AGN observations (Seaton 1978, Dumount et al.\ 1998). 

\subsection{CCF analysis}
An estimate of size and structure of the broad-line region can be obtained 
from the cross-correlation function (CCF) of a continuum light curve with
emission line light curves.\\
We cross-correlated the 5100\,\AA\ continuum light curve with all our emission
line light curves (Fig.~4) using an interpolation cross-correlation function
 method (ICCF) described by Gaskell \& Peterson (1987).
In Fig.~9 we plot the cross-correlation functions
 of the individual emission line light curves of HeII$\lambda$4686, HeI$\lambda$5876, H$\beta$ and H$\alpha$ with the continuum light curve.
\begin{figure}
%centerline{\psfig{figure=ccf4.ps,width=6.5cm,angle=270,clip=}}
\centerline{\psfig{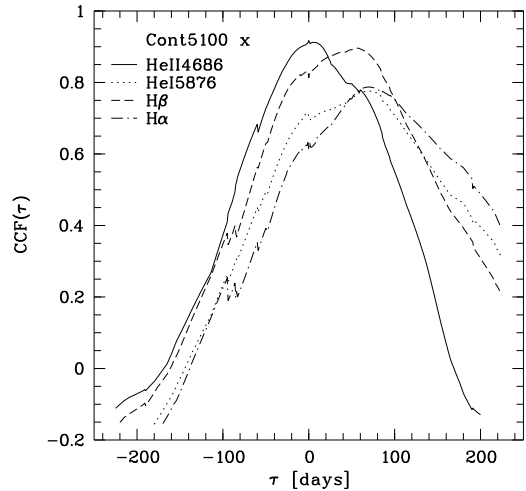}}
\caption{Cross-correlation functions CCF($\tau$) of HeII$\lambda$4686, HeI$\lambda$5876, H$\beta$ and H$\alpha$ light curves with the 5100\,\AA\ continuum light curve.}
\end{figure}
The cross-correlation functions 
 of HeI$\lambda$4471 and HeI$\lambda$5876 are identical within the errors;
% 5\%.
therefore, only one curve is shown in the plot.

First of all we determined an error of the centroids of the ICCFs by
averaging the centroids $\tau_{cent}$ that were calculated for
 fractions of the peak ranging from 35\% to 90\% of the maximum value of the
 cross-correlation functions. Then we estimated the influence of two principal
sources of cross-correlation uncertainties namely flux uncertainties
in individual measurements and uncertainties
connected to the sampling of the light curves. We used a method similar to
that described by Peterson et al.\ (1998b). 
We added random noise to our measured flux values and 
calculated the cross-correlation lags a large number of times. 
Due to the large variability amplitudes of Mkn~110 these uncertainties were of lower weight compared to those introduced by the sampling of the light curves. The sampling uncertainties were estimated by considering different subsets of our light curves and repeating the cross-correlation calculations. 
Typically we excluded
37\% of our spectra from the data set (cf.\ Peterson et al.\ 1998b). In Table 5
we list our final cross-correlation results together with the total error.

\begin{table}
\caption{Cross-Correlation Lags}
\begin{tabular}{lc}
\hline 
\noalign{\smallskip}
Line & \multicolumn{1}{c}{$\tau_{cent}$} \\
     & \multicolumn{1}{c}{[days]}\\
(1)  & \multicolumn{1}{c}{(2)}\\
\noalign{\smallskip}
\hline
\noalign{\smallskip}
HeII$\lambda 4686$~ &   $~9.4^{+~5.6}_{-12.8}$\\[.7ex]
H$\beta$            &   $39.9^{+33.2}_{-~9.5}$\\[.7ex]
HeI$\lambda 5876$~  &   $59.6^{+43.1}_{-36.9}$\\[.7ex]
HeI$\lambda 4471$   &   $62.7^{+46.5}_{-47.7}$\\[.7ex]
H$\alpha$           &   $81.6^{+29.4}_{-31.1}$\\
\noalign{\smallskip}
\hline 
\end{tabular}
\end{table}

Considering the entire observing period we got a lag
 of the H$\beta$ light curve of $39.9^{+33.2}_{-9.5}$ days.
 Peterson et al.\ (1998a) obtained for a similar extended 
 observing campaign a lag of $31.6^{+9.0}_{-7.3}$ days.
However, they claim that their best lag estimate derived from an observing
period of 123 days yielding the smallest error is about $19.5^{+6.5}_{-6.8}$ days.
%On the other hand it has been shown (e.g. Dietrich \& Kollatschny 1995) that the centroid values of the CCF depend on the light curve intervals used for the CCF calculation. 
%They might differ by e.g. ten days if one is using light curves of different years but for the same galaxy and for the same emission lines.

\subsection{Line profiles and their variations}
Normalized mean and rms profiles of HeII$\lambda$4686,
 HeI$\lambda$5876, H$\beta$ and H$\alpha$ lines are 
shown in Figs.~10 and 11.
\begin{figure}
%centerline{\psfig{figure=mean4.ps,width=6.5cm,angle=270,clip=}}
\centerline{\psfig{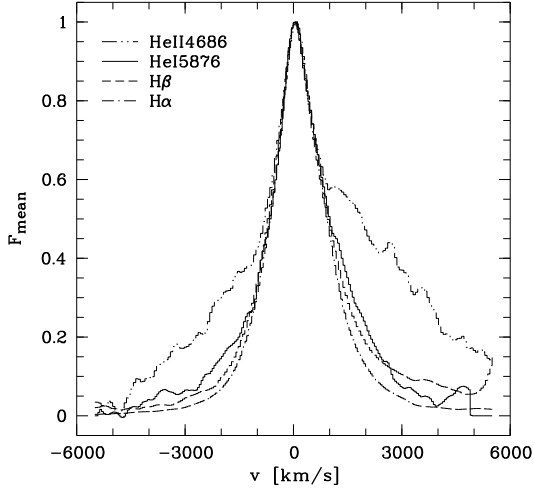}}
\caption{Mean profiles of HeII$\lambda$4686, HeI$\lambda$5876,
H$\beta$ and H$\alpha$.}
\end{figure}
\begin{figure}
%centerline{\psfig{figure=rms4.ps,width=6.5cm,angle=270,clip=}}
\centerline{\psfig{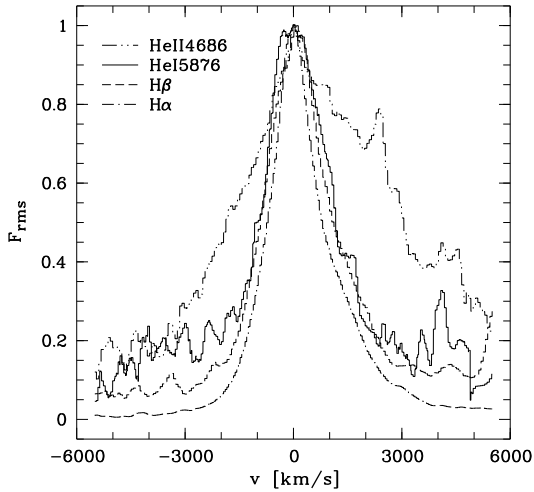}}
\caption{The rms profiles of HeII$\lambda$4686, HeI$\lambda$5876,
H$\beta$ and H$\alpha$.}
\end{figure}
The rms profile is a measure of the variable part in the line profile.
There is a very broad line component in the mean and rms profiles
especially to be seen in the HeII line.
Even apart from this very broad component the mean and rms profiles of
the individual lines are different with respect to their shape and
full width at half maximum (FWHM).
In Table~6 we list the widths 
of the mean and rms profiles. The mean and rms H$\beta$ profiles are
more similar to the HeI$\lambda$4471 profiles than to H$\alpha$.
 The rms profile of H$\alpha$ e.g.
is significantly narrower than the profile of H$\beta$.
The profiles of the HeI$\lambda$4471 line are more noisy than
the other ones. They are identical to those of the HeI$\lambda$5876 line
within the errors.

\begin{table}
\caption{Mean and rms line widths (FWHM)}
\begin{tabular}{lcc}
\hline 
\noalign{\smallskip}
Line & FWHM(mean)  & FWHM(rms) \\
     & [km s$^{-1}$] & [km s$^{-1}$] \\
(1) & (2) & (3)  \\ 
\noalign{\smallskip}
\hline
\noalign{\smallskip}
HeII$\lambda 4686$~ & 2720 $\pm$ 100 & 4930 $\pm$ 200\\
H$\beta$            & 1670 $\pm$  50 & 2010 $\pm$ 100\\
HeI$\lambda 5876$~  & 1640 $\pm$  50 & 2200 $\pm$ 100\\
H$\alpha$           & 1580 $\pm$  50 & 1540 $\pm$ 100 \\
\noalign{\smallskip}
\hline 
\end{tabular}
\end{table}

All mean and rms profiles
 show a red asymmetry. The asymmetry is mainly caused by a
second line component at v=1200 km s$^{-1}$. This second component
does not vary with the same amplitude as the main component. Furthermore,
this second component was stronger during the first half of our campaign 
from 1987 until January 1992 than during the second half of the campaign.
The H$\alpha$ spectra taken at the intensity minima of February 1989 and August 1994 are plotted in Fig.~12.
The additional component centered at v=1200 km s$^{-1}$ is clearly
to be seen.
\begin{figure}
%centerline{\psfig{figure=d89-93.ps,width=6.5cm,angle=270,clip=}}
\centerline{\psfig{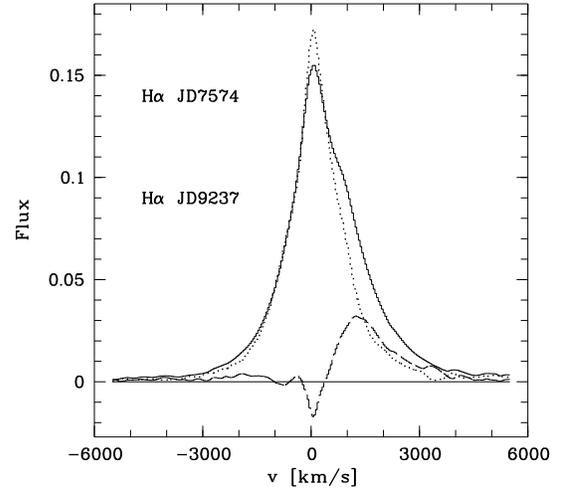}}
\caption{H$\alpha$ spectra taken at the first minimum state
 of February 1989 (solid line) as well as at 
the second minimum state of August 1994 (dotted line)
 and their difference spectrum (dashed line).}
\end{figure}
The mean spectra of the first half of our campaign are broader by
400~-~500 km s$^{-1}$ (FWHM) than those of the second half
because of this component.

There is an independent very broad component
present in the mean and rms HeII profiles
(Figs.~10 and 11). This very broad component
exists in addition to the broad component. There is
no transition component visible in the profile. The peak
of this very broad profile
component is redshifted by 400$\pm$100 km s$^{-1}$ with
 respect to the narrow lines. This shift was measured in the difference spectra (cf.\ Fig.~13). 
This very broad component is the strongest contributor 
to the HeII variability as can be seen from the rms profile.
The very broad component is visible in the Balmer line profiles also,
especially at high continuum stages (see Fig.~3). 
The HeII and H$\beta$ profiles taken in January 1992
 are shown in more detail in Fig.~13. We subtracted the minimum profile 
taken in October 1988 to remove the narrow line component.
\begin{figure}
%centerline{\psfig{figure=vblr.ps,width=6.5cm,angle=270,clip=}}
\centerline{\psfig{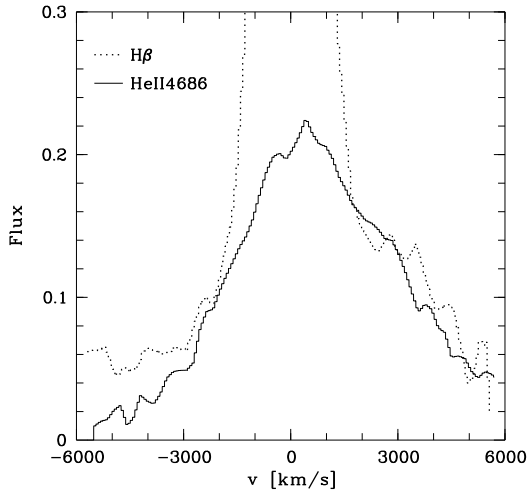}}
\caption{The very broad components of the HeII and H$\beta$ profiles
taken in January 1992 after subtraction of our minimum profile taken
in October 1988.} 
\end{figure}
The HeII line intensity has been divided by a factor of 1.3
 for direct comparison with the very broad H$\beta$ profile.
Immediately one can see the striking similarity. The blue wing of the H$\beta$
profile is stronger than that of the HeII profile
 because of the blending with the
red wing of the HeII line.
 The very broad line component
has a full width at zero intensity (FWZI) of 12\,000 km s$^{-1}$.

\section{Discussion}
Mkn~110 is one of the very few Seyfert galaxies with spectral
variability coverage over a time interval of ten years.
Different continuum ranges
 show different variability amplitudes; this holds for
 different optical emission lines, too.
But the mean fluxes of the continuum and of all emission
lines remain nearly constant integrated
over time scales of a few years (see Fig.~4).
There are considerable variations over
time scales of days to years. The strongest variability amplitudes
in the continuum shows the blue spectral range (see Figs.~2 to 4).
There are intensity variations of a factor of $>5$. The strongest
amplitudes in the blue spectral range  
might be explained by a greater share of the non-thermal continuum
with respect to the underlying galaxy continuum.

The optical line variations of H$\beta$ are very strong in comparison 
to other Seyfert galaxies (e.g. Peterson et al.\ 1998a).
The HeII$\lambda$4686 line shows the strongest
 variations of nearly a factor      
of 8 within two years. On the other hand the H$\beta$ and the
continuum($\lambda$5100) vary only by a factor of 1.7  and 3.0
respectively within the same time  interval. Apart from the
variation of the HeII$\lambda$4686 line in NGC~5548 in 1984
(Peterson \& Ferland 1986) these are the strongest optical line variations within such a 
time interval. In the case of Mkn~110 we can show that the appearance of
the very broad HeII$\lambda$4686 and H$\beta$
component (see Figs.~3 and 13) is not a unique event in the accretion rate.
It is connected to a very strong 
ionizing continuum flux as can be seen from the light curves.
The intensity ratio HeII$\lambda$4686/H$\beta$ comes to a value 
of 1.3 (see Fig.13) in the very broad line region.
 Such a line ratio is still in correspondence with
photoionization of broad emission-line clouds in
quasars (Korista et al.\ 1997).

The very broad line region (VBLR) originates close to the central
ionizing source at a distance of about 9 light days. It is
not connected to the ``normal'' BLR. As can be seen 
from the line profiles there exists no continuous transition region
between these BLRs. The center of the VBLR line profiles is shifted by 
400$\pm$100 km~s$^{-1}$ with respect to the ``normal''
 BLR profiles (Figs.\ 10, 11, 13).

Apart from this VBLR component we could show that the line profiles of the
Balmer and HeI lines are similar but not identical. The 
H$\alpha$ line profile is narrower than the H$\beta$ profile.
Besides the cross-correlation results this is an independent
indication that these two lines do not originate in exactly the same region. 

The observed Ly$\alpha$/H$\beta$ ratio comes to
a value of 11.0 in Mkn~110. This is
about a factor of two higher than the mean observed Ly$\alpha$/H$\beta$ ratio
in Seyfert galaxies (Wu et al.\ 1983). Photoionization models of
Kwan \& Krolik (1981) result in  Ly$\alpha$/H$\beta \simeq 10$ without the
presence of dust. Therefore, dust may not play an important role in
the BLR of Mkn~110.
 The Balmer decrement in Mkn~110 varies as a function of the  
ionizing continuum flux. This might be explained by
radiative transfer effects rather than by variation of dust extinction.

The profiles of the broad emission lines in Mkn~110
 are neither symmetric nor
smooth (Figs.~10, 11). This is a further indication that the broad-line regions
in AGN are structured as e.g. in NGC~4593 (Kollatschny \& Dietrich 1997).
In Fig.~12 it is shown that during the first half of our campaign a
red line component was present in the H$\alpha$ spectra. This component
was not visible during the second half of the campaign.

The size of the H$\beta$ line emitting region (r~=~40 ld corresponding to
 1.0 10$^{17}$ cm) and the optical continuum luminosity is compared to
those of other Seyfert galaxies. The continuum luminosity amounts to
$L_{5100} = 4.4~10^{39}$ erg s$^{-1}$\,\AA$^{-1}$.
In this case we used $H_{0}=100$ km s$^{-1}$ Mpc$^{-1}$ in order 
to compare directly the radius and luminosity 
of Mkn~110 with those of other Seyfert galaxies compiled by Carone et al.\ (1996). The values of Mkn~110 fit nicely into the 
general radius-luminosity relationship for the broad-line regions in
Seyfert galaxies. The Balmer line emitting region as well as
the luminosity of Mkn~110 are arranged in the upper region of the
radius-luminosity plane close to the galaxies Mkn~590 and Mkn~335.

There is a trend that the broader emission lines originate closer to the
center (see Table 7). A similar trend was found for NGC~5548, too (Kollatschny 
\& Dietrich 1996). 

The central mass in Mkn~110 can be estimated from
the width of the broad emission line profiles (FWHM) under the assumption 
that the gas dynamics are dominated by the central massive object.
Furthermore, one needs the distance of the dominant emission line clouds
to the ionizing central source
(e.g. Koratkar \& Gaskell 1991, Kollatschny \& Dietrich 1997).
We presume that the characteristic velocity of the emission line
region is given by the FWHM of the rms profile and the characteristic 
distance R is given by the centroid of the corresponding cross-correlation
 function:
\[ M = \frac{3}{2} v^{2} G^{-1} R .\] 
In Table 7 we list our virial mass estimations of the central massive object
 in Mkn~110.
\begin{table}
\caption{Virial mass estimations}
\begin{tabular}{lccc}
\hline 
\noalign{\smallskip}
Line & FWHM(rms)     & R    & M                \\
     & [km s$^{-1}$] & [ld] & [$10^7 M_{\odot}$]  \\
(1) & (2) & (3) & (4) \\ 
\noalign{\smallskip}
\hline 
\noalign{\smallskip}
HeII$\lambda 4686$~ & 4930 $\pm$ 200 & $~9.4^{+~5.6}_{-12.8}$ & $6.7^{+4.0}_{-9.1}$\\[.7ex]
HeI$\lambda 5876$~  & 2200 $\pm$ 100 & $59.6^{+43.1}_{-36.9}$ & $8.5^{+6.2}_{-5.3}$\\[.7ex]
H$\beta$            & 2010 $\pm$ 100 & $39.9^{+33.2}_{-~9.5}$ & $4.7^{+4.0}_{-1.2}$\\[.7ex]
H$\alpha$           & 1540 $\pm$ 100 & $81.6^{+29.4}_{-31.1}$ & $5.7^{+2.2}_{-2.3}$\\
\noalign{\smallskip}
\hline 
\end{tabular}
\end{table}
Altogether we determine a central mass of:
\[ M= 6.4^{+2.2}_{-2.7}\, 10^{7} M_{\odot} . \]

We can independently estimate an upper limit of the central mass
if we interpret the observed redshift of the very broad HeII component
($\Delta z = 0.0013$) %(v = 400 km~s$^{-1}$)
as gravitational redshift (e.g.~Zheng \& Sulentic 1990):
\[ M = c^{2} G^{-1} R \Delta z \]
Again we presume that this line 
component originates at a distance of 9 ld from the central ionizing source.
We derive an upper limit of the central mass of
\[ M = 2.1~10^{8} M_{\odot}  \]
This second independent method confirms the former mass estimation.

\section{Summary}
Mkn~110 shows strong variations in the continuum and in the line intensities
on time scales of days to years.
The continuum - especially the blue range - varies by a factor of
3 to 5 on time scales of years. The Balmer line intensities vary by a factor 
of 2.5 while the HeII$\lambda$4686 line shows exceptionally strong variations by a factor of 8.

We cross-correlated the light curves of the emission lines with those of
the continuum. The emission lines originate at distances of 9 to 80
light days from the central source as a function of ionization degree.

Not only the line intensities but also the line profiles varied. We
detected a very broad line region VBLR  component in the high intensity stages
of the Balmer and HeII lines. This region exists separated from the
``normal'' broad-line region at a distance of only 9 light days 
from the central ionizing source.

We derived the central mass %of $M=(6.4^{2.2}_{2.7}) 10^{7} M_{\odot} $
in Mkn~110 using two independent methods.

Mkn~110 is a prime target for further detailed variability studies
with respect to the line and continuum variability amplitudes as well as
with respect to the short-term variations.

\begin{acknowledgements}
We thank M.\ Dietrich, D.\ Grupe, and U.\ Thiele for taking spectra for us. We are grateful to M.\ Dietrich, E.\ van Groningen, and I.\ Wanders who made available some software to us. 
This work has been supported by DARA grant 50~OR\-9408\-9 and DFG grant Ko 857/13.
\end{acknowledgements}

\newpage
\begin{landscape}
\setcounter{table}{1}
\begin{table}
\caption{Continuum and integrated line fluxes}
\begin{tabular}{cccccccccc}
\noalign{\smallskip}
\hline 
\noalign{\smallskip}
Julian Date & 5100\,\AA & 4265\,\AA & 3750\,\AA & H$\alpha$ & H$\beta$ & HeII$\lambda 4686$ & HeI$\lambda 4471$ & HeI$\lambda 5876$ \\
2\,440\,000+\\
(1) & (2) & (3) & (4) & (5) & (6) & (7) & (8) & (9) \\ 
\noalign{\smallskip}
\hline
\noalign{\smallskip}
6828&    3.14  $\pm\ $0.09  &   3.97  $\pm\ $0.1   &    9.0  $\pm\ $0.2   &   1656  $\pm\ $24    &   477.9 $\pm\ $ 9.8  &   108.3 $\pm\ $5.2   &   30.8  $\pm\ $5.2   &   68.7  $\pm\ $5.4   &   \\
7229&    2.57  $\pm\ $0.08  &   2.83  $\pm\ $0.1   &    5.2  $\pm\ $0.2   &   1499  $\pm\ $22    &   389.3 $\pm\ $ 8.9  &    58.1 $\pm\ $4.2   &   20.6  $\pm\ $4.4   &   54.0  $\pm\ $4.7   &   \\   
7438&    1.77  $\pm\ $0.07  &   1.96  $\pm\ $0.1   &    2.9  $\pm\ $0.4   &   1112  $\pm\ $18    &   287.3 $\pm\ $ 7.9  &    37.2 $\pm\ $3.7   &   17.6  $\pm\ $4.2   &   33.1  $\pm\ $3.7   &   \\   
7574&    3.06  $\pm\ $0.09  &   3.53  $\pm\ $0.1   &    6.6  $\pm\ $0.4   &   1439  $\pm\ $21    &   347.0 $\pm\ $ 8.5  &    60.5 $\pm\ $4.2   &   26.7  $\pm\ $4.9   &   51.1  $\pm\ $4.6   &   \\   
7663&    4.43  $\pm\ $0.12  &   5.27  $\pm\ $0.1   &   12.5  $\pm\ $0.4   &   1744  $\pm\ $24    &   475.1 $\pm\ $ 9.8  &   159.2 $\pm\ $6.2   &   25.8  $\pm\ $4.8   &   61.8  $\pm\ $5.1   &   \\   
7828&    1.98  $\pm\ $0.07  &   1.99  $\pm\ $0.1   &                   &   1520  $\pm\ $22    &   357.9 $\pm\ $ 8.6  &    30.6 $\pm\ $3.6   &   22.5  $\pm\ $4.6   &   41.7  $\pm\ $4.1   &   \\   
8632&    5.93  $\pm\ $0.15  &   7.56  $\pm\ $0.1   &   15.1  $\pm\ $0.4   &   1995  $\pm\ $27    &   624.5 $\pm\ $11.2  &   237.6 $\pm\ $7.8   &   41.7  $\pm\ $5.9   &   103.  $\pm\ $7.2   &   \\   
8812&    3.24  $\pm\ $0.09  &                   &                   &   2091  $\pm\ $28    &   494.9 $\pm\ $ 9.9  &   101.2 $\pm\ $5.0   &   36.1  $\pm\ $5.5   &   95.6  $\pm\ $6.8   &   \\   
8860&    2.86  $\pm\ $0.09  &                   &                   &   1673  $\pm\ $24    &   421.4 $\pm\ $ 9.2  &    80.7 $\pm\ $4.6   &   35.4  $\pm\ $5.5   &   70.7  $\pm\ $5.5   &   \\   
8862&    3.26  $\pm\ $0.10  &                   &                   &   1782  $\pm\ $25    &   454.9 $\pm\ $ 9.5  &    92.1 $\pm\ $4.8   &   32.6  $\pm\ $5.3   &   81.7  $\pm\ $6.1   &   \\   
8864&    3.30  $\pm\ $0.10  &   3.00  $\pm\ $0.2   &    4.8  $\pm\ $0.2   &   1703  $\pm\ $24    &   432.1 $\pm\ $ 9.3  &    80.7 $\pm\ $4.6   &   32.0  $\pm\ $5.2   &   79.7  $\pm\ $6.0   &   \\   
9078&    3.15  $\pm\ $0.09  &                   &                   &   1692  $\pm\ $24    &   468.5 $\pm\ $ 9.7  &    91.2 $\pm\ $4.8   &   30.9  $\pm\ $5.2   &   80.4  $\pm\ $6.0   &   \\   
9080&    3.12  $\pm\ $0.09  &                   &                   &   1702  $\pm\ $24    &   453.9 $\pm\ $ 9.5  &    88.2 $\pm\ $4.8   &   23.9  $\pm\ $4.7   &   80.4  $\pm\ $6.0   &   \\   
9083&    3.22  $\pm\ $0.09  &                   &                   &   1764  $\pm\ $25    &   467.7 $\pm\ $ 9.7  &    86.6 $\pm\ $4.7   &   27.8  $\pm\ $4.9   &   77.6  $\pm\ $5.9   &   \\   
9123&    3.47  $\pm\ $0.10  &                   &                   &   1678  $\pm\ $24    &   482.3 $\pm\ $ 9.8  &   159.4 $\pm\ $6.2   &   40.9  $\pm\ $5.9   &   87.3  $\pm\ $6.4   &   \\   
9237&    1.82  $\pm\ $0.07  &   2.01  $\pm\ $0.2   &    4.1  $\pm\ $0.4   &   1197  $\pm\ $19    &   310.2 $\pm\ $ 8.1  &    65.5 $\pm\ $4.3   &   25.6  $\pm\ $4.8   &   59.0  $\pm\ $4.9   &   \\   
9419 &    1.91  $\pm\ $0.07  &                   &                   &                   &   288.0 $\pm\ $ 7.9  &    81.9 $\pm\ $4.6   &                   &                   &   \\   
9595&    1.53  $\pm\ $0.06  &                   &                   &    876  $\pm\ $16    &   266.8 $\pm\ $ 7.7  &    46.5 $\pm\ $3.9   &   12.0  $\pm\ $3.8   &   38.4  $\pm\ $3.9   &   \\   
9776&    4.00  $\pm\ $0.11  &   4.95  $\pm\ $0.1   &                   &   1432  $\pm\ $21    &   414.6 $\pm\ $ 9.1  &   133.9 $\pm\ $5.7   &   29.1  $\pm\ $5.0   &   68.1  $\pm\ $5.4   &   \\   
9785&    4.17  $\pm\ $0.11  &   4.57  $\pm\ $0.1   &                   &   1430  $\pm\ $21    &   411.8 $\pm\ $ 9.1  &   135.9 $\pm\ $5.7   &   27.9  $\pm\ $5.0   &   72.5  $\pm\ $5.6   &   \\   
9786&    3.95  $\pm\ $0.11  &   4.30  $\pm\ $0.1   &                   &   1600  $\pm\ $23    &   412.9 $\pm\ $ 9.1  &   128.4 $\pm\ $5.6   &   22.6  $\pm\ $4.6   &   74.9  $\pm\ $5.7   &   \\   
9787&    3.96  $\pm\ $0.11  &   4.44  $\pm\ $0.1   &                   &   1529  $\pm\ $22    &   433.6 $\pm\ $ 9.3  &   128.2 $\pm\ $5.6   &   30.9  $\pm\ $5.2   &   72.1  $\pm\ $5.6   &   \\   
9811&    3.62  $\pm\ $0.10  &   4.36  $\pm\ $0.1   &                   &   1485  $\pm\ $22    &   441.1 $\pm\ $ 9.4  &    95.1 $\pm\ $4.9   &   35.0  $\pm\ $5.4   &   69.3  $\pm\ $5.5   &   \\   
9870&    4.86  $\pm\ $0.13  &                   &                   &   1558  $\pm\ $23    &   445.9 $\pm\ $ 9.5  &   174.2 $\pm\ $6.5   &                      &   75.8  $\pm\ $5.8   &   \\   
\noalign{\smallskip}
\hline 
\noalign{\smallskip}
\end{tabular}

Continuum fluxes (2) - (4) in 10$^{-15}$\,erg\,sec$^{-1}$\,cm$^{-2}$\,\AA$^{-1}$.\\
Line fluxes (5) - (9) in 10$^{-15}$\,erg\,sec$^{-1}$\,cm$^{-2}$.
\end{table}
\end{landscape}
\end{document}